\documentclass[aps,pra,twocolumn,showpacs,showkeys]{revtex4}
\usepackage{graphicx}
\usepackage{dcolumn}
\usepackage{amssymb} 
 
\usepackage{amsfonts} 
\usepackage{amssymb} 
\usepackage{amsmath} 
\usepackage{color}
\usepackage{hyperref}
 
\newcommand{\ar}{\arrowvert}

\newcommand{\ov}{\overline} 
\newcommand{\cd}{\! \cdot \!} 
\newcommand{\be}{\begin{equation}} 
\newcommand{\ee}{\end{equation}} 
\newcommand{\ba}{\begin{eqnarray}} 
\newcommand{\ea}{\end{eqnarray}}

\newcommand{\nn}{\nonumber}

\newcommand{\pa}{\partial}

\begin{document} 
\title{Shear and bulk viscosities of a photon gas at low temperature}
\author{Antonio Dobado$^1$, Felipe J. Llanes-Estrada$^1$ and Juan M. Torres-Rincon$^2$}
\affiliation{ 
$^1$ Departamento de F\'{\i}sica Te\'orica I,  Universidad 
Complutense, 28040 Madrid, Spain. \\
$^2$ Institut de Ci\`encies de l'Espai (IEEC/CSIC), Campus Universitat Aut\`onoma de Barcelona, 
Facultat de Ci\`encies, Torre C5, E-08193 Bellaterra, Spain
} 
 
\date{\today}

\begin{abstract}  
We explore the viscosities of a photon gas by means of the Euler-Heisenberg effective theory and quantum electrodynamics at zero electron chemical potential. We find parametric estimates that show a
very large shear viscosity and an extremely small bulk viscosity (reflecting the very weak coupling simultaneously with a very approximate dilatation invariance). The system is of some interest because it exemplifies very neatly the influence
of the breaking of scale invariance on the bulk viscosity.
\end{abstract} 
\maketitle

\section{Introduction} \label{sec:intro}

RHIC experiments have drawn much attention to the quark and gluon plasma as a nearly perfect fluid~\cite{Romatschke:2007mq}. The outcome of an extensive body of investigations is that, near the phase transition to a hadron medium, the fluid is very strongly coupled, featuring a minimum in the shear viscosity~\cite{Csernai:2006zz,Dobado:2008vt} (normalized to the entropy density). Under current discussion is the behavior of the bulk viscosity~\cite{Karsch:2007jc}
and whether~\cite{Dobado:2012zf,FernandezFraile:2008vu} or not~\cite{FernandezFraile:2010gu} there is a maximum of the bulk viscosity in the same crossover region.  Also well known are the transport coefficients in perturbative quantum chromodynamics (QCD)~\cite{Arnold:2003zc}, as well as the
low-energy pion gas~\cite{Lu:2011df,Dobado:2011qu} and strongly coupled Fermi systems~\cite{Mannarelli:2012su}.

In this brief report we examine quite an opposite example: a photon gas at a very low temperature. 
Classical electrodynamics is a linear theory, in which light beams cross each other without interacting. Thus, all transport is effected by interactions with the cavity walls such as in a waveguide, and one cannot really talk of a fluid in infinite matter. 

However in quantum electrodynamics (QED), a photon can fluctuate instantly into an electron-positron pair, and another photon can Compton-scatter off the virtual charge thus created. Therefore, QED is no more a linear theory, and photon-photon scattering is possible. This physics is becoming of wider interest due to the advent of high intensity laser fields~\cite{PVLAS}, that may allow studies of the phenomenon. Meanwhile, the only classical experimental example is probably the well-known Delbr\"uck scattering.

A case in point for an infinite photon gas is the cosmic microwave background (CMB) at $T=2.72$ K~\cite{Ade:2011ap}; 
since the photon gas is so weakly coupled at such low temperatures, photon mean free paths are astronomic and thus the shear viscosity (a purely diffusive phenomenon) is
also huge. Our parametric estimate is given in Eq.~(\ref{shear}) in Sec.~\ref{sec:summary} as a function of the temperature and electron mass.

The bulk viscosity turns out to be a tiny coefficient due to the nearly scale invariance of the photon gas.
As both elastic and inelastic scattering can occur, then two possibilities appear. Hence, in Sec.~\ref{sec:summary} we provide two different parametric results.

If the system is long-lived, one needs to keep the slowest relaxing mode in the inversion of the linearized Boltzmann collision operator in Eq.~(\ref{Boltzmann2}) below. 
Then inelastic processes clearly dominate the bulk viscosity, chemical equilibration is achieved and the chemical potential must be zero. In a gas of on-shell quasiphotons, the lowest order kernel that achieves this is $\gamma\gamma\to 4\gamma$.
The source function in the left-hand side of the Boltzmann equation~(\ref{Boltzmann2}), is orthogonal to the unique zero mode in the collision operator (the energy zero mode, when $A(p)$ is proportional to the energy).
The Boltzmann equation is then compatible~\cite{Arnold:2003zc}.

On the other hand, one is often interested in not-so-long times (or the system under consideration has a short duration), understood as those that are short compared
with the chemical equilibration time given by that inelastic reaction. Then, the kinetic equilibration time is given in the photon gas by the elastic reactions
$\gamma\gamma\to \gamma\gamma$. An effective chemical potential could be introduced and additional terms appear in the source function (see discussion and the extra terms
for a massive pion gas in Ref.~\cite{Dobado:2011qu}). Additionally to the energy zero mode in the collision term, an extra
zero mode related to the particle conservation --when $A(p)$ in Eq.~(\ref{Boltzmann2}) is a constant-- is also present. However, this mode is also orthogonal to the source function.
Thus, the Boltzmann equation is again compatible and its inversion can then be carried out.

The two parametric estimates for the bulk viscosity are given in Eqs.~(\ref{bulkinelastic}) and~(\ref{bulkelastic}), for the case with dominant inelastic/elastic interaction, respectively.

The effective Lagrangian of Euler-Heisenberg~\cite{Heisenberg:1935qt} is not sufficient for these estimates of the bulk viscosity due to thermal non-analyticities. The
effective theory is constructed as a derivative power series and by standard counting, $T\sim p$ so that one expects observables to be organized in terms of $(T/m_e)^n$,
where $m_e$ is the electron mass. The exponentials in Eq.~(\ref{bulkinelastic}) and~(\ref{bulkelastic}) are singularities not captured in such framework.
This is due to the need of breaking scale invariance with a thermal photon mass, which is not possible in a one-loop perturbative calculation in the effective theory.
Thus, we turn to a full calculation of the thermal photon mass within QED, and thus the bulk viscosity, in Sec.~\ref{sec:QED}, where happily we can draw much from the
literature. Our conclusions and outlook are presented in Sec.~\ref{sec:summary}.

\section{Photon-photon scattering in the Euler-Heisenberg effective theory} \label{sec:gammagamma}

\begin{figure}[h]
\begin{center}
\includegraphics[width=0.35\textwidth]{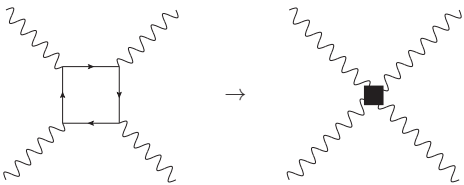}
\end{center}
\caption{\label{fig:scattering} Left: Box diagram for photon-photon scattering in QED. Right: Effective contact interaction in photon-photon scattering at low energies. }
\end{figure}

For low-frequency radiation, the box diagram (left panel of Fig.~\ref{fig:scattering}) necessary to calculate the photon-photon scattering amplitude in QED reduces to the effective Lagrangian density of Euler-Heisenberg, expressible in a gauge-invariant way in terms of the electric and magnetic fields as
\be
\mathcal{L}_{EH} = \frac{1}{2}({\bf E}^2-{\bf B}^2) +
\frac{e^4}{360\pi^2m_e^4} \left[ ({\bf E}^2-{\bf B}^2)^2+
7({\bf E}\cd{\bf B})^2\right] \ .
\ee

The Feynman rule for the photon-photon vertex (depicted in the right panel of the same Fig.~\ref{fig:scattering}) is extracted from the interaction part of this Lagrangian, that, when written as a function of the electromagnetic tensor, reads
\be \label{EulerHeisenberg}
\mathcal{L}= a F_{\alpha \beta} F^{\alpha \beta} F_{\mu \nu} F^{\mu \nu} + b F_{\alpha \beta} F_{\mu \nu} F^{\alpha \mu} F^{\beta \nu}\ , \ee
with low-energy  constants $a = -\frac{\alpha^2}{36 m_e^4}$ and $b = \frac{7\alpha^2}{90 m_e^4}$ and $\alpha \simeq 1/137$ the electromagnetic coupling constant.

The photon four-point function at tree level $M_{\alpha \beta \mu \nu} (p_1,p_2,p_3,p)$ was obtained not too long ago in~\cite{Halter:1993kj}. 
Although we only use in this brief note its parametric dependence, since we have detected some typos and to provide all necessary information should the numeric coefficient of order one multiplying the viscosities ever be needed, we provide it again in App.~\ref{appendix}.

A handy nontrivial check --since one wrong sign would make it fail -- is the explicit satisfaction of the gauge Ward identities, viz.
\ba p_1^{\alpha} M_{\alpha \beta \mu \nu} (p_1,p_2,p_3,p)&= & 0 \ , \\
 p_2^{\beta} M_{\alpha \beta \mu \nu}(p_1,p_2,p_3,p) &=& 0 \ , \\
 p_3^{\mu} M_{\alpha \beta \mu \nu}(p_1,p_2,p_3,p) &=& 0 \ , \\
 p^{\nu} M_{\alpha \beta \mu \nu}(p_1,p_2,p_3,p) &=& 0 \ . \ea
 
The photon-photon scattering amplitude with momentum assignments $p_1^{\alpha} p_2^{\beta} \rightarrow p_3^{\mu} p^{\nu}$ is then 
\be \label{4amplitude}
i {\cal M}=i\epsilon_{\alpha}^{(1)} \epsilon_\beta^{(2)} \epsilon^{*(3)}_\mu \epsilon^{*(4)}_\nu M^{\alpha \beta \mu \nu} \ . \ee

Squaring and averaging over initial polarizations, and summing over final polarizations, then using the Ward identities, and introducing Mandelstam variables for the on-shell photons (all of them standard manipulations), we  obtain
\be \label{ampsquare}
\ov{|{\cal M}|^2}= \frac{139}{2025} \frac{\alpha^4}{m_e^8} \left( t^4 + 2 st^3 + 3 s^2t^2 + 2 s^3 t +s^4\right) \ . \ee

This result, so obtained from the low-energy theory, coincides exactly with the computation in Ref.~\cite{Liang:2011sj}, using QED at one loop (the box diagram in Fig.~\ref{fig:scattering}), and only later taking the low-energy limit.
(This scattering amplitude is needed in the collision integral of the Boltzmann equation if the numeric coefficient of the viscosities is to be evaluated: we do not see a motivation to perform such detailed numerical calculation at the present time, but details on the corresponding Chapman-Enskog expansion can be found in~\cite{Dobado:2008vt}).

To check known results once more, we take Eq.~(\ref{ampsquare}) to the center of mass frame
\be \ov{|{\cal M}|^2}= \frac{139}{2025} \frac{\alpha^4}{m_e^8} 16 \omega^8  \left( \cos^2 \theta +3 \right)^2  \ , \ee
where $\omega$ in the center-of-mass energy, and then calculate the differential cross section,
\be \frac{d \sigma_{\gamma \gamma \rightarrow \gamma \gamma}}{d \Omega_{CM}} = \frac{1}{2} \frac{1}{64 \pi^2} \frac{1}{(2\omega)^2}\ov{|{\cal M}|^2} \ , \ee
(where a factor $1/2$ has been included because the two photons are identical in the final state).
The result
\be 
\frac{d \sigma_{\gamma \gamma \rightarrow \gamma \gamma}}{d \Omega_{CM}} = \frac{139}{64800\pi^2} \frac{\alpha^4 \omega^6}{m_e^8}  \left( \cos^2 \theta +3 \right)^2  
\ee
can be integrated to obtain the total cross section~\cite{Karplus:1950zza,Liang:2011sj}
\be \label{eq:crosssec} \sigma_{\gamma \gamma \rightarrow \gamma \gamma} =  \frac{973}{10125\pi} \frac{\alpha^4 \omega^6}{m_e^8}  \ . \ee

The photons are very nearly free. Thus, quasiparticle kinetic theory is a very accurate starting point. The shear viscosity of the photon gas is then expressible as an integral over the shearing function $\delta f_p \propto B(p)$ (all details are given e.g. in~\cite{Torres-Rincon:2012sda}) that characterizes
the separation from the equilibrium Bose-Einstein function $f_p\equiv n_p$ in the Landau-Lifschitz reference frame,
\be \label{shear2}
\eta = \frac{2}{15 T^3} \int \frac{d^3p}{(2\pi)^3E_p} n_p (1+n_p) p^6 B(p)\ .
\ee

Parametrizing a small separation from equilibrium, $B(p)$ satisfies a linearized Boltzmann-like equation (Uehling-Uhlenbeck equation)
\ba \label{Boltzmann}
n_p (1+n_p) p^i p^j =  \frac{1}{2T^2} \int (1+n_1)(1+n_2) n_3 n_p
\nonumber \\  \nonumber
  \times \ov{\ar {\mathcal M}\ar^2} \prod_{k=1}^3 \frac{d^3p_k}{2 E_k (2\pi)^3}
(2\pi)^4 \delta^4(p_1+p_2-p_3-p) \\
 \times \left( p^i p^j B(p)\!+\!p_3^i p_3^j B(p_3)\!-\! p_1^i p_1^j B(p_1)\!-\! p_2^i p_2^j B(p_2)
\right)\ .
\ea

A full numerical evaluation is beyond our present scope. Nevertheless, the parametric dependence of the shear viscosity can already be obtained by examining these two equations, or by employing the relaxation-time approximation with
a thermally-averaged cross section~(\ref{eq:crosssec}):
\be \eta \sim \frac{T}{\ov{\sigma}_{\gamma\gamma\to\gamma\gamma}} \sim \frac{1}{\alpha^4} \frac{m_e^8}{T^5} \ . \ee
This establishes Eq.~(\ref{shear}) below. We now proceed to the bulk viscosity.

\section{Bulk viscosity and thermal non-analyticity} \label{sec:QED}

\subsection{Bulk viscosity in the effective theory} \label{subsec:effective}

An evaluation of the bulk viscosity along the lines of Sec.~\ref{sec:gammagamma} is bound to fail. To see it, write down the equivalent of Eqs.~(\ref{shear2}) and (\ref{Boltzmann}), where the disturbance from equilibrium is $\delta f_p \propto A(p)$:
\be \label{bulk}
\zeta = \frac{2}{T} \int \frac{d^3p}{(2\pi)^3E_p} n_p (1+n_p) A(p) \frac{E_p 
(\mathbf{p} \cdot \mathbf{\nabla}_p) E_p}{3} 
\ee
and
\ba \label{Boltzmann2}
n_p(1+n_p) \left(
\frac{\mathbf{p} \cdot \mathbf{\nabla}_p E_p}{3}  - v_s^2 \frac{\pa (\beta E_p)}{\pa \beta} 
\right) =  
\nonumber \\
\int (1+n_1) (1+n_2) n_3 n_p \left[ A (p) + A(p_3)-A(p_1) - A(p_2) \right]  \nonumber \\ 
\times \frac{1}{2E_p} \  \ov{\ar {\mathcal M}\ar^2} \prod_{k=1}^3 \frac{d^3p_k}{2 E_k (2\pi)^3}
(2\pi)^4 \delta^4(p_1+p_2-p_3-p) \ , \nonumber \\
\ea
where $\beta=1/T$ and $v_s^2$ denotes the adiabatic speed of sound squared.

In a dilatation-invariant theory, such as Maxwell's is, $\zeta$ vanishes directly because the factor 
$\left( \frac{\mathbf{p} \cdot \mathbf{\nabla}_p E_p}{3}  - v_s^2 \frac{\pa (\beta E_p)}{\pa \beta} 
\right)$ 
averages to zero upon integrating Eq.~(\ref{Boltzmann2}) over $p$. 
The electron mass, that breaks dilatation invariance, does enter the constants $a$ and $b$ in the low-energy theory Eq.~(\ref{EulerHeisenberg}), that appear in the squared, average amplitude $\ov{\ar {\mathcal M}\ar^2}$ in Eq.~(\ref{Boltzmann2}), but as we now examine, does not affect the vanishing of the factor in question in the low-energy effective theory,  \emph{because this is controlled by the quasiparticle (photon) mass}, not by the electron mass. 

Indeed, the one-loop correction to the photon vacuum polarization using the Euler-Heisenberg theory was calculated
 in Ref.~\cite{Thoma:2000fd}. It leads to a dispersion relation for tranverse photons at low $T$ that is linear in $p$,
\be \label{disp_rel} E_p = \sqrt{\frac{1-\gamma}{1+\gamma}} \ p \ , \ee
with the speed of light reduced by
\be 
 \gamma = \frac{44 \pi^2}{2025} \frac{\alpha^2 T^4}{m_e^4} \ . 
\ee
Therefore, the low-energy theory does not generate a thermal mass for the photon. In fact, this one-loop masslessness is valid at any order in the perturbative expansion~\cite{Andersen:2001kt}.

To see the vanishing of the bulk viscosity  in a bit more detail, it is convenient to reduce Eq.~(\ref{bulk}) to
\be \label{bulk2} 
\zeta = \frac{2}{T} \frac{1-\gamma}{1+\gamma} \int \frac{d^3p}{(2\pi)^3E_p} n_p (1+n_p) A(p) \frac{p^2}{3} \ .
\ee
But in the Landau-Lifschitz formalism, in addition to the Landau-Lifschitz condition that fixes the reference frame, 
there is a so-called ``condition of fit'' that fixes the energy content of the system. 
It involves the $00$-component of the stress-energy tensor~\cite{Jeon:1995zm},
\be  \label{condfit} 
0\equiv \tau^{00} =  \int  \frac{-2d^3p}{(2\pi)^3 E_p} n_p (1+n_p) A(p)  
\ E_p \frac{\pa (\beta E_p)}{\pa \beta}  \ . 
\ee
Using the dispersion relation in~(\ref{disp_rel}), \emph{without a photon thermal mass}, this yields
\be  \label{condfit2}
\int  \frac{d^3p}{(2\pi)^3 E_p} n_p (1+n_p) A(p)  p^2 =0 \ . \ee
Since the bulk viscosity in Eq.~(\ref{bulk2}) is proportional to the integral in Eq.~(\ref{condfit2}), it must vanish.
This happens even as the speed of sound differs from the conformal value $v_s^2- 1/3=- 4 \gamma$ because each of the terms in $\left( \frac{\mathbf{p} \cdot \mathbf{\nabla}_p E_p}{3}  - v_s^2 \frac{\pa (\beta E_p)}{\pa \beta} 
\right)$ separately yields a zero integral!

\subsection{One-loop QED}

We have just found that the bulk viscosity is predicted to vanish within the Euler-Heisenberg effective Lagrangian at the lowest order. This is a surprising feature, since the Lagrangian is supposed to represent photon-photon interactions in QED at that low momentum and zero electron chemical potential (that is, in practice, at zero electron density).

We now show that this is not the case in QED, and that the bulk viscosity is actually calculably finite (although, at CMB temperatures, it is tiny). 
In one-loop QED, the dispersion relation has been calculated~\cite{Nieves:1983fk,Kapusta:2006pm,Andersen:2001kt}
\be  \label{dis_rel2} 
E_p^2=p^2+m^2_\gamma \ , 
\ee
with a photon thermal mass
\be m_\gamma^2= \frac{8\alpha }{\sqrt{2\pi}}  m_e^{3/2} T^{1/2} e^{-m_e/T} \ . \ee
We immediately see the difficulty with the effective theory. This thermal mass contains the factor 
$e^{-m_e/T}$ that admits no Taylor expansion around $T=0$. Therefore, the usual counting $p\sim T$ does not lead to a polynomial behavior upon expanding. Since in QED we do know the microscopic theory, we can proceed with the mass in Eq.~(\ref{dis_rel2}) and repeat the reasoning of subsection~\ref{subsec:effective}.

We use the condition of fit~(\ref{condfit}) to introduce an extra term in the bulk viscosity Eq.~(\ref{bulk}) so that the integrand coincides with the left-hand side of the Boltzmann equation~(\ref{Boltzmann2}).  This is
\ba \label{bulk3}
\zeta &=& \frac{2}{T} \int \frac{d^3p}{(2\pi)^3 E_p} n_p (1+n_p) A(p) \nonumber \\ 
 &\times &\left[ E_p^2 \left( \frac{1}{3} - v_s^2\right) - \frac{m^2_\gamma}{3} +\frac{v_s^2 T}{2} \frac{\pa m^2_\gamma }{\pa T} \right] \ ,
\ea
where Eq.~(\ref{dis_rel2}) has been used.

To reduce this equation further we need to calculate the speed of sound $v_s$. 
We thus turn to the thermodynamics of a noninteracting (ideal) quasiparticle gas~\cite{lifshitz1980statistical} following the same 
methodology of QCD exposed at length in~\cite{Gorenstein:1995vm,Arnold:2003zc,Khvorostukhin:2010cw}.

The entropy density of an ideal photon gas with dispersion relation $E_p^2=p^2+m^2_\gamma$ is, to order $\alpha$
\be \label{eq:entropy} s(T)=\frac{4\pi^2}{45} T^3 - \frac{4\alpha}{3\sqrt{\pi}} m_e^{3/2} T^{3/2} e^{-m_e/T} \ . \ee
From this expression we can compute the speed of sound:
\ba 
 v_s^2 = \frac{s(T)}{T\frac{ds(T)}{dT}} &= &\frac{1}{3} - \frac{ 5 \sqrt{2}}{24 \pi^2} \frac{m_e}{T} \frac{m^2_\gamma}{T^2}  \nonumber \\ 
      &=& \frac{1}{3} - \frac{5\alpha}{3 \pi^{5/2}} \frac{m_e^{5/2}}{T^{5/2}} e^{-m_e/T} \ .
\ea

At $\mathcal{O} (\alpha)$, the source function in the bulk viscosity, 
the last bracket in Eq.~(\ref{bulk3}) reads
\ba\label{nonconformal}
\left[
E_p^2 \left( \frac{1}{3} - v_s^2\right) - \frac{m^2_\gamma}{3} + \frac{v_s^2 T}{2} \frac{\pa m^2_\gamma }{\pa T}
\right] \nonumber \\
= \
m^2_\gamma \left( \frac{ 5 \sqrt{2}}{24 \pi^2} \frac{m_e}{T} \frac{E_p^2}{T^2} - \frac{1}{4} \right) 
\ea
which is proportional to the square of the photon thermal mass, 
and therefore nonzero for one-loop QED at finite temperature. 

If only the parametric dependence is desired, straightforward algebra starting in Eq.~(\ref{bulk3}), combined with Eq.~(\ref{Boltzmann2}) and Eq.~(\ref{nonconformal}), leads to 
\be 
\zeta \sim \frac{1}{\alpha^4} \frac{m_e^{10}}{T^{10}} \frac{m_\gamma^4}{T} \ , 
\ee
that can also be obtained from the simple estimate,
\be 
\zeta \sim \frac{T}{\ov{\sigma}_{\gamma\gamma\to\gamma\gamma}} \left( \frac{1}{3}-v_s^2 \right)^2
\ee
and immediately yields Eq.~(\ref{bulkelastic}) in Sec.~\ref{sec:summary}.

Returning now to infinitely slow relaxation, Furry's theorem forbids a finite five-point function. 
The six-point function is suppressed respect to Eq.~(\ref{4amplitude}) by $e^2/m_e^2$. This is because of the two additional photons attached, and the consequent two additional fermion propagators.
Squaring  this suppression factor, we immediately obtain the estimate in Eq.~(\ref{bulkinelastic}) below.

\section{Summary and discussion} \label{sec:summary}

Both shear and bulk viscosity have attracted much attention in low-energy hadron physics recently, where the methods of effective theories are widely used. 

We present parametric estimates of these coefficients in Eqs.~(\ref{shear}),(\ref{bulkinelastic}) and~(\ref{bulkelastic}), and provided enough detail to undertake a
numerical evaluation via a Chapman-Enskog expansion of Boltzmann's equation, should the need ever arise. 

\begin{figure}[h]
\begin{center}
\includegraphics[width=0.4\textwidth]{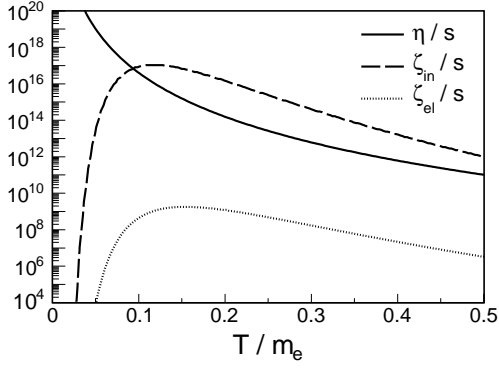}
\end{center}
\caption{\label{fig:viscosities} Shear viscosity over entropy density (solid line) and the bulk viscosity over entropy density due to inelastic (dashed) and elastic (dotted) processes. The numerical coefficients $\eta_0=\zeta_0=\zeta'_0$ are set to one.}
\end{figure}

The shear viscosity is huge ($\eta_0$ being the numerical coefficient not computed):
\be \label{shear}
\eta = \eta_0 \frac{1}{\alpha^4} \frac{m_e^8}{T^5} \ \simeq \ \eta_0 \frac{(1.4\cd 10^{19} \rm{eV})^3}{T^5(\rm{K})} \ ,
\ee
which is typical of such weakly coupled boson gases~\cite{Manuel:2004iv}.
The estimate follows trivially from the extensive discussion above in Sec.~\ref{sec:gammagamma} where we give enough detail as to calculate the numeric coefficient if it was ever necessary.
Normalized to the entropy density in Eq.~(\ref{eq:entropy}) the coefficient reads at leading order:
\be 
\frac{\eta}{s} \propto \frac{1}{\alpha^4} \frac{m_e^8}{T^8} \ .
\ee
We show this coefficient as a function of temperature (using $\eta_0=1$) in Fig.~\ref{fig:viscosities}.

The bulk viscosity, on the contrary, is tiny in infinite matter at temperatures much below the electron mass,
\be \label{bulkinelastic}
\zeta = \zeta_0 \frac{1}{\alpha^4} \frac{m_e^{17}}{T^{14}} e^{-2m_e/T} \ ,
\ee
in the bona-fide hydrodynamic limit for arbitrarily long-wave modes out of equilibrium, a regime where inelastic $\gamma\gamma \to 4 \gamma$ processes dominate the relaxation.
$\zeta_0$ is the constant coefficient for the bulk viscosity dominated by inelastic processes.
Normalized to the entropy density the bulk viscosity reads:
\be
\frac{\zeta}{s} \propto \frac{1}{\alpha^4} \frac{m_e^{17}}{T^{17}} e^{-2m_e/T} \ . 
\ee
If one considers times that are not so large, so that the modes relaxing are purely kinetic, the bulk viscosity is then dominated in effect by elastic photon-photon scattering 
\be \label{bulkelastic}
\zeta = \zeta'_0 \frac{1}{\alpha^2} \frac{m_e^{13}}{T^{10}} e^{-2m_e/T}\ ,
\ee
with $\zeta'_0$ the numerical coefficient independent of $m_e$ and $T$. Normalized to the entropy density
\be 
\frac{\zeta}{s} \propto \frac{1}{\alpha^2} \frac{m_e^{13}}{T^{13}} e^{-2m_e/T}\ .
\ee
Both bulk viscosities are tiny when compared to the shear viscosity.
For example, evaluating Eq.~(\ref{bulkelastic}) we find $\zeta \sim T^{-10}({\rm K}) \cd e^{2(127.3-m_e/T)} {\rm eV}^3$. Since $m_e/T\simeq 2.2\cd 10^9$ for the CMB temperature today, the exponent is huge and negative, so that the bulk viscosity is incredibly small (it of course gets bigger as we proceed back in time towards recombination).
This smallness is because the photon gas is very nearly a relativistic, conformally invariant gas, whose bulk viscosity is well-known to be zero~\cite{Weinberg}.

When the temperature becomes comparable, or if an electron chemical potential is introduced (finite electron density) then the population of electrons dominates the physics and the situation is very different, which we hope to address in a future publication. Compton (eventually Thomson) scattering slows down diffusive momentum transfer at tree level, instead of the one loop in QED necessary if only photons are present.

The huge ratio $\eta/\zeta$ is typical of very weakly coupled systems, and quite the opposite of what is being seen at RHIC, where, near the phase transition, the bulk viscosity could actually play an important role. 
In the photon gas, the bulk viscosity turns out to be so small because, in spite of the weak (polynomial) coupling, the breaking of conformal invariance is exponentially suppressed. To see this we have resorted to the microscopic QED theory, 
since the effective theory of Euler-Heisenberg for the photons alone seems to miss the photon thermal mass due to its non-analytic behavior.


\begin{appendix}
\section{Photon four-point function\label{appendix}}
We provide the corrected four-point function at tree level from the Euler-Heisenberg Lagrangian~(\ref{EulerHeisenberg}) (to be compared with the result of Ref.~\cite{Halter:1993kj}):
 \begin{widetext}
\ba \nn iM^{\alpha \beta \mu \nu} (p_1,p_2,p_3,p)   = \\
 \nn 	g^{\alpha \beta} g^{\mu \nu} (-32ia p_1 \cdot p_2 p_3 \cdot p - 8ibp_1 \cdot p_3 p_2 \cdot p - 8i b p_1 \cdot p p_2 \cdot p_3) \\
\nn   +   g^{\alpha \mu} g^{\beta \nu} (-32ia p_1 \cdot p_3 p_2 \cdot p - 8ibp_1 \cdot p_2 p_3 \cdot p - 8i b p_1 \cdot p p_2 \cdot p_3) \\
\nn   +   g^{\alpha \nu} g^{\beta \mu} (-32ia p_1 \cdot p p_2 \cdot p_3 - 8ibp_1 \cdot p_2 p_3 \cdot p - 8i b p_1 \cdot p_3 p_2 \cdot p) \\
\nn   -  g^{\alpha \beta} \left\{ -32 i a p_3^{\nu} p^{\mu} p_1 \cdot p_2 \right.
\nn   +   8ib \left. \left[ p_3 \cdot p (p_1^{\mu} p_2^{\nu} + p_1^{\nu} p_2^{\mu}) - p_1^{\mu} p_3^{\nu} p_2 \cdot p 
- p_2^{\nu} p^{\mu} p_1 \cdot p_3 - p_2^{\mu} p_3^{\nu} p_1 \cdot p - p_1^{\nu} p^{\mu} p_2 \cdot p_3 \right] \right\} \\
\nn   -  g^{\alpha \mu} \left\{ -32 i a p_2^{\nu} p^{\beta} p_1 \cdot p_3 \right.
\nn   +   8ib \left. \left[ p_2 \cdot p (p_1^{\nu} p_3^{\beta} + p_1^{\beta} p_3^{\nu} )- p_1^{\beta} p_2^{\nu} p_3 \cdot p 
- p_3^{\nu} p^{\beta} p_1 \cdot p_2 - p_2^{\nu} p_3^{\beta} p_1 \cdot p - p_1^{\nu} p^{\beta} p_2 \cdot p_3 \right] \right\} \\
\nn   -  g^{\alpha \nu} \left\{ -32 i a p_2^{\mu} p_3^{\beta} p_1 \cdot p \right.
\nn   +   8ib \left. \left[ p_2 \cdot p_3 (p_1^{\mu} p^{\beta} + p_1^{\beta} p^{\mu} )- p_3^{\beta} p^{\mu} p_1 \cdot p_2 
- p_2^{\mu} p^{\beta} p_1 \cdot p_3 - p_1^{\mu} p_3^{\beta} p_2 \cdot p - p_1^{\beta} p_2^{\mu} p_3 \cdot p \right] \right\} \\
\nn   -  g^{\mu \nu} \left\{ -32 i a p_1^{\beta} p_2^{\alpha} p_3 \cdot p \right. 
\nn   +   8ib \left. \left[ p_1 \cdot p_2 (p_3^{\alpha} p^{\beta} + p_3^{\beta} p^{\alpha}) - p_1^{\beta} p^{\alpha} p_2 \cdot p_3 
- p_1^{\beta} p_3^{\alpha} p_2 \cdot p - p_2^{\alpha} p^{\beta} p_1 \cdot p_3 - p_2^{\alpha} p_3^{\beta} p_1 \cdot p \right] \right\} \\
\nn   -  g^{\mu \beta} \left\{ -32 i a p_1^{\nu} p^{\alpha} p_2 \cdot p_3 \right. 
\nn   +   8ib \left. \left[ p_1 \cdot p (p_2^{\nu} p_3^{\alpha} + p_2^{\alpha} p_3^{\nu}) - p_2^{\nu} p^{\alpha} p_1 \cdot p_3 
- p_3^{\nu} p^{\alpha} p_1 \cdot p_2 - p_1^{\nu} p_3^{\alpha} p_2 \cdot p - p_1^{\nu} p_2^{\alpha} p_3 \cdot p \right] \right\} \\
\nn   -  g^{\nu \beta} \left\{ -32 i a p_1^{\mu} p_3^{\alpha} p_2 \cdot p \right. 
   +   8ib \left. \left[ p_1 \cdot p_3 (p_2^{\mu} p^{\alpha} + p_2^{\alpha} p^{\mu}) - p_2^{\mu} p_3^{\alpha} p_1 \cdot p 
- p_1^{\mu} p_2^{\alpha} p_3 \cdot p - p_3^{\alpha} p^{\mu} p_1 \cdot p_2 - p_1^{\mu} p^{\alpha} p_2 \cdot p_3 \right] \right\} \\
\nn  \! -\!   32ia( p_1^{\mu} p_2^{\nu} p_3^{\alpha} p^{\beta}\!+\!p_1^{\nu} p_2^{\mu} p_3^{\beta} p^{\alpha}\!+\!p_1^{\beta} p_2^{\alpha} p_3^{\nu} p^{\mu} ) 
 \! -\!   8ib ( p_1^{\mu} p_2^{\nu} p_3^{\beta} p^{\alpha}\!+\!p_1^{\mu} p_2^{\alpha} p_3^{\nu} p^{\beta}\!+\!p_1^{\nu} p_2^{\alpha} p_3^{\beta} p^{\mu}
\!+\! p_1^{\nu} p_2^{\mu} p_3^{\alpha} p^{\beta}\!+\!p_1^{\beta} p_2^{\nu} p_3^{\alpha} p^{\mu}\!+\!p_1^{\beta} p_2^{\mu} p_3^{\nu} p^{\alpha}) \ .
\ea
\end{widetext}
\end{appendix}

\begin{acknowledgments}
We thank useful conversations with T. Schaefer. This work was supported by Spanish grants FPA2010-16963 and FPA2011-27853-C02-01. JMTR is funded by 
grant FP7-PEOPLE-2011-CIG under contract number PCIG09-GA-2011-291679. 
\end{acknowledgments}



\begin{thebibliography}{99} 

\bibitem{Romatschke:2007mq} 
  See for example P.~Romatschke and U.~Romatschke,
  Phys.\ Rev.\ Lett.\  {\bf 99}, 172301 (2007)
  [arXiv:0706.1522 [nucl-th]];
  C.~Gale, S.~Jeon and B.~Schenke,
  Int.\ J.\ Mod.\ Phys.\ A {\bf 28}, 1340011 (2013)
  [arXiv:1301.5893 [nucl-th]].

\bibitem{Csernai:2006zz} 
  L.~P.~Csernai, J.~I.~Kapusta and L.~D.~McLerran,
  Phys.\ Rev.\ Lett.\  {\bf 97}, 152303 (2006)
  [nucl-th/0604032].

\bibitem{Dobado:2008vt} 
  A.~Dobado, F.~J.~Llanes-Estrada and J.~M.~Torres-Rincon,
  Phys.\ Rev.\ D {\bf 79}, 014002 (2009)
  [arXiv:0803.3275 [hep-ph]];
  A.~Dobado, F.~J.~Llanes-Estrada and J.~M.~Torres-Rincon,
  Phys.\ Rev.\ D {\bf 80}, 114015 (2009)
  [arXiv:0907.5483 [hep-ph]].

\bibitem{Karsch:2007jc} 
  F.~Karsch, D.~Kharzeev and K.~Tuchin,
  Phys.\ Lett.\ B {\bf 663}, 217 (2008)
  [arXiv:0711.0914 [hep-ph]].

\bibitem{Dobado:2012zf} 
  A.~Dobado and J.~M.~Torres-Rincon,
  Phys.\ Rev.\ D {\bf 86}, 074021 (2012)
  [arXiv:1206.1261 [hep-ph]].


\bibitem{FernandezFraile:2008vu} 
  D.~Fernandez-Fraile and A.~Gomez Nicola,
  Phys.\ Rev.\ Lett.\  {\bf 102}, 121601 (2009)
  [arXiv:0809.4663 [hep-ph]].

\bibitem{FernandezFraile:2010gu} 
  D.~Fernandez-Fraile,
  Phys.\ Rev.\ D {\bf 83}, 065001 (2011)
  [arXiv:1009.2741 [hep-ph]].

\bibitem{Arnold:2003zc} 
  P.~B.~Arnold, G.~D.~Moore and L.~G.~Yaffe,
  JHEP {\bf 0305}, 051 (2003)
  [hep-ph/0302165];
  P.~B.~Arnold, C.~Dogan and G.~D.~Moore,
  Phys.\ Rev.\ D {\bf 74}, 085021 (2006)
  [hep-ph/0608012].
  
\bibitem{Lu:2011df} 
  E.~Lu and G.~D.~Moore,
  Phys.\ Rev.\ C {\bf 83}, 044901 (2011)
  [arXiv:1102.0017 [hep-ph]];

 \bibitem{Dobado:2011qu} 
  A.~Dobado, F.~J.~Llanes-Estrada and J.~M.~Torres-Rincon,
  Phys.\ Lett.\ B {\bf 702}, 43 (2011)
  [arXiv:1103.0735 [hep-ph]].

\bibitem{Mannarelli:2012su}
  M.~Mannarelli, C.~Manuel and L.~Tolos,
  Annals Phys. 336 (2013) 12-35
  [arXiv:1212.5152 [cond-mat.quant-gas];
  
  M.~Mannarelli, C.~Manuel and L.~Tolos,
  arXiv:1201.4006 [cond-mat.quant-gas];
  T.~Schaefer and K.~Dusling,
  arXiv:1305.4688 [cond-mat.quant-gas].

\bibitem{PVLAS}
E. Milotti {\it et al.}, Int. J. Quantum Inform. 10, 1241002 (2012) 


\bibitem{Ade:2011ap} 
  P.~A.~R.~Ade {\it et al.}  [Planck Collaboration],
  Astron.\ Astrophys.\  {\bf 536}, A18 (2011)
  [arXiv:1101.2028 [astro-ph.CO]].


 
\bibitem{Heisenberg:1935qt} 
  W.~Heisenberg and H.~Euler,
  Z.\ Phys.\  {\bf 98}, 714 (1936)
  [physics/0605038].
 
\bibitem{Halter:1993kj} 
  J.~Halter,
  Phys.\ Lett.\ B {\bf 316}, 155 (1993).

\bibitem{Liang:2011sj} 
  Y.~Liang and A.~Czarnecki,
  Can.\ J.\ Phys.\  {\bf 90}, 11 (2012)
  [arXiv:1111.6126 [hep-ph]].

    
\bibitem{Karplus:1950zza} 
  R.~Karplus and M.~Neuman,
  Phys.\ Rev.\  {\bf 80}, 380 (1950).
  
  
\bibitem{Torres-Rincon:2012sda} 
  J.~M.~Torres-Rincon,
  ``Hadronic Transport Coefficients from Effective Field Theories'' (Ph.D. Dissertation) Universidad Complutense de Madrid (Spain), 2012,
  [arXiv:1205.0782 [hep-ph]].

  
  
\bibitem{Thoma:2000fd} 
  M.~H.~Thoma,
  Europhys.\ Lett.\  {\bf 52}, 498 (2000)
  [hep-ph/0005282].



\bibitem{Andersen:2001kt} 
  J.~O.~Andersen,
  Phys.\ Rev.\ D {\bf 65}, 025014 (2002)
  [hep-ph/0103285].
  


\bibitem{Jeon:1995zm} 
  S.~Jeon and L.~G.~Yaffe,
  Phys.\ Rev.\ D {\bf 53}, 5799 (1996)
  [hep-ph/9512263];
  P.~Chakraborty and J.~I.~Kapusta,
  Phys.\ Rev.\ C {\bf 83}, 014906 (2011)
  [arXiv:1006.0257 [nucl-th]].

\bibitem{Nieves:1983fk} 
  J.~F.~Nieves, P.~B.~Pal and D.~G.~Unger,
  Phys.\ Rev.\ D {\bf 28}, 908 (1983);
  O.~K.~Kalashnikov,
  Phys.\ Scripta {\bf 58}, 310 (1998)
  [hep-ph/9802427].
  
\bibitem{Kapusta:2006pm} 
  J.~I.~Kapusta and C.~Gale,
  ``Finite-Temperature Field Theory'',
  Cambridge University Presss, UK (2006) 


\bibitem{lifshitz1980statistical}
   E.~M.~Lifschitz and L.~P.~Pitaevskii,
   ``Statistical Physics Part 2, Landau and Lifshitz Course of Theoretical Physics Vol. 9'',
   Pergamon Press, Oxford (1980)


\bibitem{Gorenstein:1995vm} 
  M.~I.~Gorenstein and S.~-N.~Yang,
  Phys.\ Rev.\ D {\bf 52}, 5206 (1995).


\bibitem{Khvorostukhin:2010cw} 
  A.~S.~Khvorostukhin, V.~D.~Toneev and D.~N.~Voskresensky,
  Phys.\ Rev.\ C {\bf 83}, 035204 (2011)
  [arXiv:1011.0839 [nucl-th]].


\bibitem{Manuel:2004iv} 
  C.~Manuel, A.~Dobado and F.~J.~Llanes-Estrada,
  JHEP {\bf 0509}, 076 (2005)
  [hep-ph/0406058].



\bibitem{Weinberg}
 S.~Weinberg, 
 ``Gravitation and cosmology: Principle and applications of general theory of relativity'',
  IE-Wiley, New York  (1972).


\end{thebibliography}
\end{document}